\documentclass[10pt,journal,compsoc]{IEEEtran}
\usepackage{microtype}
\ifCLASSINFOpdf
   \usepackage[pdftex]{graphicx}
   \graphicspath{./figures/}
   \DeclareGraphicsExtensions{.pdf,.jpeg,.png}
\else
   \usepackage[dvips]{graphicx}
   \graphicspath{{../eps/}}
   \DeclareGraphicsExtensions{.eps}
\fi
\usepackage{multirow}
\usepackage{subfigure}
\usepackage{booktabs} 
\usepackage{mathtools}
\usepackage{hyperref}
\usepackage{algorithm}
\usepackage[noend]{algpseudocode}
\algdef{SE}[DOWHILE]{Do}{doWhile}{\algorithmicdo}[1]{\algorithmicwhile\ #1}%
\usepackage{amsmath}

\ifCLASSOPTIONcompsoc
  \usepackage[nocompress]{cite}
\else
  \usepackage{cite}
\fi

\begin{document}

\title{Software-Defined Design Space Exploration for an Efficient DNN~Accelerator Architecture}

\author{Ye~Yu,
        Yingmin~Li,
        Shuai~Che,
        Niraj~K.~Jha,~\IEEEmembership{Fellow,~IEEE}
        and~Weifeng~Zhang
\IEEEcompsocitemizethanks{\IEEEcompsocthanksitem Y. Yu and N. K. Jha are with the Department
of Electrical Engineering, Princeton University, Princeton,
NJ, 08544.
E-mail: \{yeyu, jha\}@princeton.edu.
Y. Li, S. Che and W. Zhang are with Alibaba Group, Sunnyvale, CA, 94085.
E-mail: \{yingmin.li, shuai.che, weifeng.z\}@alibaba-inc.com.
This work was funded by an Alibaba summer internship and NSF under Grant No. CCF-1811109.}}


\IEEEtitleabstractindextext{%
\begin{abstract}
Deep neural networks (DNNs) have been shown to outperform conventional machine learning algorithms 
across a wide range of applications, e.g., image recognition, object detection, robotics, and 
natural language processing. However, the high computational complexity of DNNs often necessitates 
extremely fast and efficient hardware. The problem gets worse as the size of neural networks 
grows exponentially. As a result, customized hardware accelerators have been developed to accelerate 
DNN processing without sacrificing model accuracy. However, previous accelerator design studies have 
not fully considered the characteristics of the target applications, which may lead to sub-optimal 
architecture designs. On the other hand, new DNN models have been developed for better accuracy, but 
their compatibility with the underlying hardware accelerator is often overlooked. In this article, we 
propose an application-driven framework for architectural design space exploration of DNN 
accelerators. This framework is based on a hardware analytical model of individual DNN
operations. It models the accelerator design task as a multi-dimensional optimization problem. We 
demonstrate that it can be efficaciously used in application-driven accelerator architecture design: 
we use the framework to optimize the accelerator configurations for eight representative DNNs and 
select the configuration with the highest geometric mean performance. The geometric mean performance 
improvement of the selected DNN configuration relative to the architectural configuration optimized 
only for each individual DNN ranges from 12.0\% to 117.9\%. Given a target DNN, the framework can 
generate efficient accelerator design solutions with optimized performance and area. Furthermore, we 
explore the opportunity to use the framework for accelerator configuration optimization under 
simultaneous diverse DNN applications.  The framework is also capable of improving neural network 
models to best fit the underlying hardware resources. We demonstrate that it can be used to analyze 
the relationship between the operations of the target DNNs and the corresponding accelerator 
configurations, based on which the DNNs can be tuned for better processing efficiency on the given 
accelerator without sacrificing accuracy. 
\end{abstract}

}

\maketitle

\IEEEdisplaynontitleabstractindextext

\IEEEpeerreviewmaketitle

\ifCLASSOPTIONcompsoc
\IEEEraisesectionheading{\section{Introduction}\label{intro}}
\else
\section{Introduction}\label{intro}
\fi

\IEEEPARstart{D}{eep} neural networks (DNNs) have recently become the fundamental inference vehicle 
for a broad range of artificial intelligence applications. Unlike conventional machine learning 
algorithms that rely on handcrafted features, DNNs extract features and learn the hidden patterns 
automatically from the training data. However, the superior performance of DNNs comes at the cost of 
requiring an immense amount of training data and massive computational complexity.
The number of operations in the winning DNNs in the ImageNet Large Scale
Visual Recognition Competition \cite{ILSVRC15} has increased exponentially over the past
few years, e.g., 1.4 GOPs for AlexNet \cite{krizhevsky2012imagenet} in 2012 to 
38 GOPs for VGG-19 \cite{VGG} in 2014. On the other hand, the slowdown of Moore's Law scaling makes 
it difficult to keep pace with DNN growth, since the gap between the computational demand from DNNs 
and the computational capacity available from underlying hardware resources keeps increasing. Hence, 
specialized hardware architectures have been developed to efficiently process DNNs. DNNs 
map well to a graphical processing unit (GPU) due to its parallel architecture, massive number
of computational units, and high memory bandwidth. However, GPU's performance density, the computation 
capacity per unit area (floating-point operations per second/mm$^{2}$), has almost saturated since 2011 \cite{xu2018scaling}. The only improvement is 
due to technology scaling from 28 nm to 20 nm \cite{xu2018scaling}. It is infeasible to continuously 
increase the chip area to accommodate larger DNNs without improving performance density.

Customized application-specific integrated circuits (ASICs)- and field-programmable gate array 
(FPGA)-based accelerators have also emerged for efficient DNN processing. These customized 
accelerators are designed to accelerate common low-level DNN operations, such as convolution and 
matrix multiplication. Hence, even though new DNN models are evolving rapidly and differ in their 
network architectures significantly, ASIC- and FPGA-based accelerators are still capable of processing 
various DNNs efficiently. FPGA-based accelerators \cite{Suda,Qiu,Han,Zhang,Zhao,underutilization,logical,level,Kwon} provide high 
parallelism and fast time-to-market. For example, an embedded FPGA platform is used as a convolver with 
dynamic-precision data quantization in \cite{Qiu} to achieve high throughput. In \cite{Han}, a 
load-balance-aware pruning method is implemented on an FPGA to compress
the long short-term memory (LSTM)
model by 20$\times$. A dynamic programming algorithm is used to map DNNs to a deeply pipelined FPGA 
in \cite{Zhang}. It can achieve up to 21$\times$ and 2$\times$ energy efficiency relative to 
a central processing unit (CPU) and GPU, respectively. To avoid the long latency of the external memory access, an FPGA-based DNN accelerator is developed in \cite{only} that keeps all the weigths and activations in the on-chip buffer. Alternatively, ASIC-based accelerators 
\cite{EIE,Venkataramani,dadiannao,TPU,flex,sparsenn,scnn,ucnn,cnvlutin,cambriconx} demonstrate much better power efficiency for DNN processing 
relative to general-purpose processors. For example, a model compression method is utilized in \cite{EIE}, where 
the inference engine can process the compressed network model efficiently through acceleration on 
sparse matrix-vector multiplications. A custom multi-chip machine-learning architecture, called 
DaDianNao, is presented in \cite{dadiannao}. With each chip implemented as a single-instruction 
multiple-data (SIMD)-like processor, DaDianNao achieves two orders of magnitude speedup over a GPU. 
In \cite{Venkataramani}, an ASIC accelerator is proposed for DNN training on the server end. It uses 
heterogeneous processing tiles with a low-overhead synchronization mechanism. A Fast Fourier Transform-based fast convolution is used in \cite{circnn} to speed up convolutional layers, where convolutions are converted into matrix multiplications. Multiple convolutional layers are fused and processed in \cite{fuse} to keep the intermediate data in the on-chip buffer, avoiding accesses to the external memory. The Google tensor 
processing unit (TPU) speeds up DNN inference with its multiply-accumulate (MAC) units 
arranged in a systolic structure \cite{TPU}. It has extended support for DNN training in its 
second version. A row-stationary scheme, a dataflow used to minimize data movement, is proposed in \cite{eye} for a spatial accelerator architecture.

To improve accelerator performance density, the computational resources should be fully utilized 
using an efficient dataflow. Since convolutional layers typically constitute over 90\% of total 
DNN operations \cite{krizhevsky2012imagenet,rtl}, parallelizing convolution computations 
can accelerate overall DNN processing significantly. The design space for convolutional layer 
processing comprises processing of multiple loops and data partitioning choices governed by 
limited on-chip memory \cite{Ma}. In order to efficiently process the convolutional layer, loop 
unrolling, loop tiling, and loop interchange are often used in recent DNN accelerators 
\cite{Zhangfpga,Ma}. Compared to fast DNN model evolution, hardware accelerator implementations are much slower. Several systematic design space exploration methods have been proposed in order to bridge this gap \cite{poster,builder,nnest}. However, previous accelerator designs do not fully consider the target 
applications in the early design stage. This may lead to the choice of a sub-optimal design from 
the DNN accelerator design space for the target applications since the characteristics of different 
DNNs may vary significantly and require very different architectural designs for efficient 
processing. 

In this article, we make the following contributions:
\newline
\indent1) We develop an application-driven framework for architectural design space exploration of 
efficient DNN accelerators. This framework is based on a hardware analytical model for various DNN 
operations. The design space exploration task is modeled as a multi-dimensional optimization problem, 
which can be handled using a genetic algorithm.
\newline
\indent2) We use this framework in the early accelerator architecture design stage to achieve 
geometric mean performance improvements ranging from 12.0\% to 117.9\%.
\newline
\indent3) We use this framework to explore optimization opportunities for simultaneously
addressing diverse DNN applications.
\newline
\indent4) We perform a sensitivity study of the relationship between DNN characteristics and the 
corresponding accelerator design configurations. 

The rest of the article is organized as follows. Section~\ref{background} discusses background 
information on DNNs. Section~\ref{kernel} presents our hardware analytical model for DNN operations. 
Section~\ref{framework} describes the application-driven framework for efficient accelerator design 
space exploration. Section~\ref{results} presents experimental results obtained by our framework on 
accelerator optimization and sensitivity study of DNN characteristics. Section~\ref{discussion} discusses the limitations of our framework and describes some potential future work. Section~\ref{conclusion} 
concludes the article.

\begin{figure*}[!t]
\centering
\includegraphics[width=6.75in]{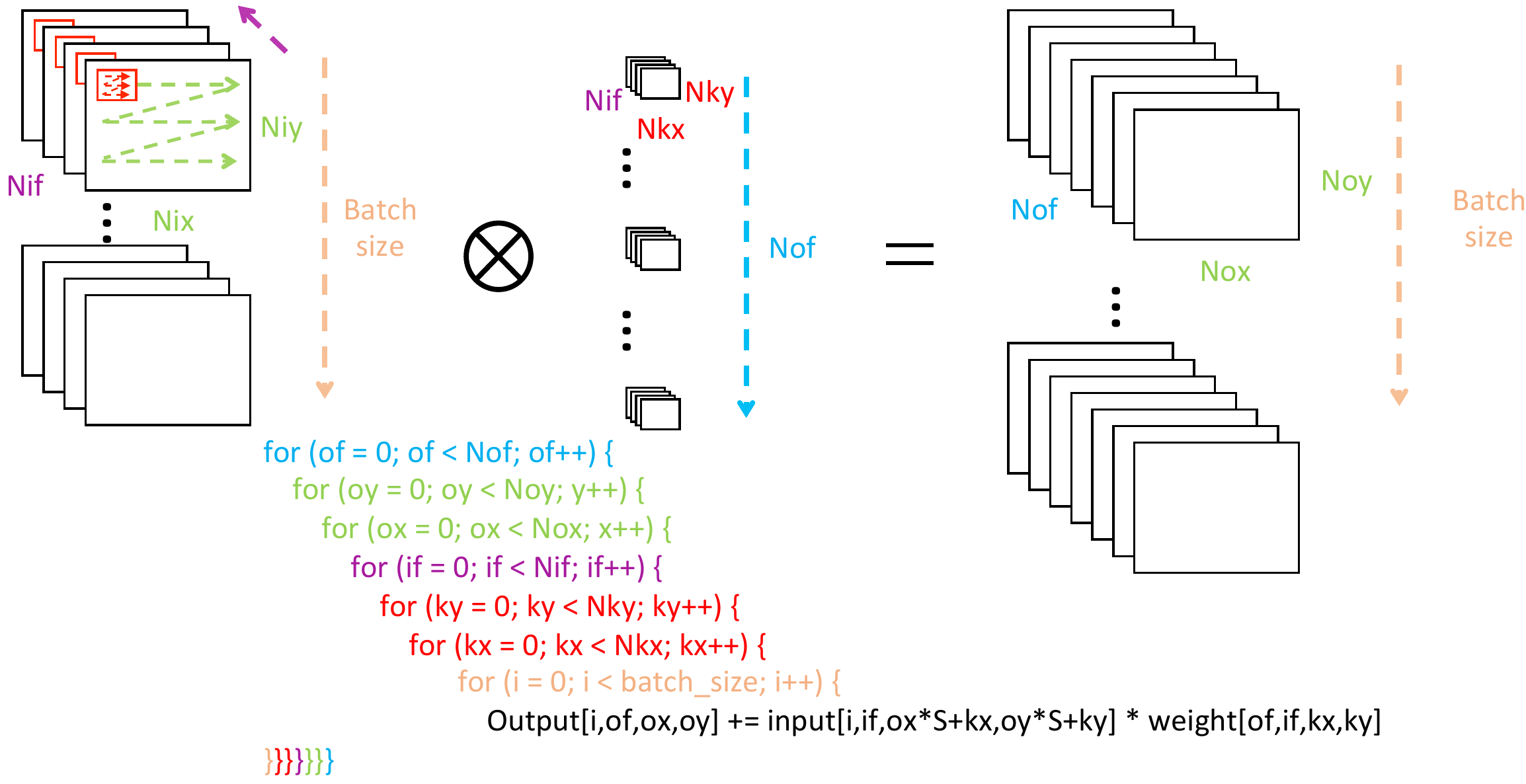}
\caption{Convolutional layer of a DNN \cite{Ma}}
\label{convolution}
\end{figure*}

\section{Background}\label{background}
In this section, we present background material to help understand the rest of the article. We 
first describe the convolutional layer of a DNN in Section \ref{conv}. We then explore two 
optimization strategies used for DNNs: computational parallelism and data reuse, in Section 
\ref{parallel} and Section \ref{reuse}, respectively. 

\begin{figure*}[!t]
\centering
\includegraphics[width=6.75in]{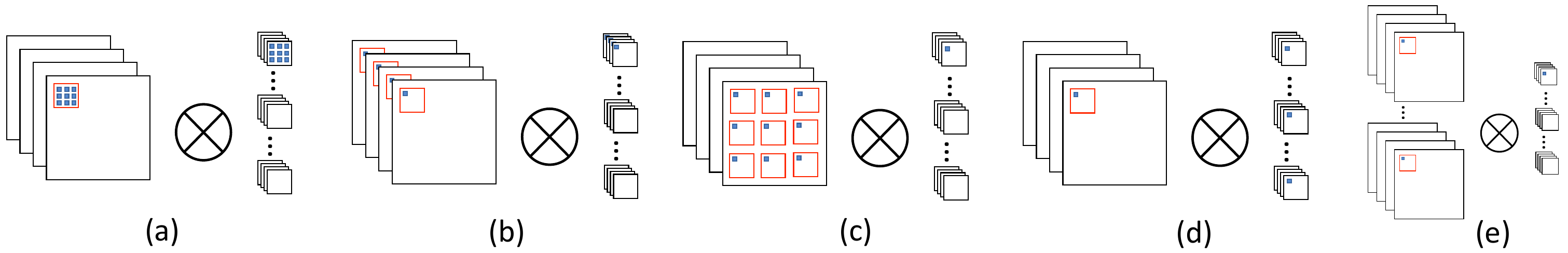}
\caption{Loop unrolling \cite{Ma}: (a)~within one kernel window, (b)~across input feature map 
channels, (c)~within one input feature map, (d)~across output feature map channels, and (e)~across 
inputs in a batch}
\label{unrolling}
\end{figure*}

\subsection{Convolutional layer of a DNN} \label{conv}
The convolutional layer is one of the DNN building blocks. In the forward pass, the convolutional 
layer convolves a batch of 3D input feature maps with multiple learnable 3D kernel weights to 
generate a batch of 3D output feature maps. These learnable 3D kernel weights are trained to detect 
which features, such as edges, are present in the input feature maps. They also capture spatial 
and temporal dependencies in the input feature maps. As shown in Fig.~\ref{convolution}, the 
convolutional layer operations can be represented by five sets of convolution loops: looping through 
different kernels, looping within one input feature map channel, looping through multiple input 
feature map channels, looping within one kernel channel, and looping through different inputs in the 
batch. The execution order of these loops can be interchanged to optimize the dataflow. N$of$, N$oy$, 
N$ox$, N$if$, N$ky$, N$kx$, batch\_size, and S denote the number of output feature map channels, 
height and width of the output feature map, number of input feature map channels, height and width 
of the kernel window, input batch size, and sliding stride, respectively. These parameters are 
determined by the architecture of the DNNs. The on-chip memory of the accelerator may not be large 
enough to hold the entire data set for input feature maps, kernel weights, and output feature maps. 
Therefore, they need to be partitioned into smaller data chunks in order to fit in the on-chip 
memory. This is called loop tiling \cite{Zhangfpga}. T* parameters (e.g., T$of$) denote the 
corresponding design variables for loop tiling. They represent the size of the data chunks stored in 
on-chip memory. Computational parallelism in the convolutional layer comes from loop unrolling, 
which increases the acceleration throughput and resource utilization ratio. P* parameters denote the 
number of parallel computations in each dimension. As the computing resources can only process 
data stored in an on-chip memory and the tiled data set is a subset of the total data set, the design 
constraints are set as P*~$\leq$~T*~$\leq$~N*. For instance, P$if\leq$~T$if\leq$~N$if$. The 
computational parallelism enabled by loop unrolling is discussed next.

\subsection{Computational parallelism}\label{parallel}
Fig.~\ref{unrolling}(a)-(e) depict how the five types of loop unrolling work in the convolutional 
layer. Fig.~\ref{unrolling}(a) depicts loop unrolling within one kernel window. In each cycle, 
P$kx\times$P$ky$ input pixels and kernel weights in the kernel window are read in from the on-chip 
memory to perform parallel multiplications followed by P$kx\times$P$ky-1$ additions. The result 
is then added to the previously obtained partial sum. However, as the kernel sizes (N$kx$ and N$ky$) 
are usually relatively small, stand-alone loop unrolling within one kernel window cannot provide 
enough parallelism to fully utilize the accelerator compute resources \cite{space}. 
Fig.~\ref{unrolling}(b) depicts loop unrolling across multiple input feature map channels. In each 
cycle, pixels at the same position in different channels are multiplied with the corresponding 
kernel weights across channels to produce P$if$ products. Then these products are summed up and 
accumulated with the partial sum. Fig.~\ref{unrolling}(c) shows loop unrolling within one input 
feature map channel.  P$ix\times$P$iy$ input pixels are multiplied with the same kernel weight in 
parallel. The products are then accumulated to the corresponding partial sums in parallel. The 
input feature map sizes (N$ix$ and N$iy$) are typically large enough to provide sufficient 
parallelism for the accelerator as long as the required data are stored in the on-chip memory. 
Fig.~\ref{unrolling}(d) describes loop unrolling across different kernels. Multiple kernel weights 
at the same location from different kernels are multiplied with the same input pixels. The products 
are accumulated to the corresponding partial sums for different output feature map channels. 
Fig.~\ref{unrolling}(e) presents loop unrolling across inputs in the batch. Multiple inputs can be 
processed in parallel since there is no data dependency among them. These loop unrolling types can 
be combined to further increase the parallelism in convolutional layer processing. For example, 
loop unrolling within the kernel window, across multiple input feature map channels, and across 
different kernels are employed together in \cite{Qiu,scale,space} while loop unrolling within one 
kernel window and within one input feature map channel are utilized in \cite{eye}.

\subsection{Data reuse}\label{reuse}
A convolutional layer is processed by sliding the kernel windows along the 3D input feature maps 
where MAC operations are performed at each sliding step. Since memory access and data movement 
incur significant delay and energy overheads \cite{eye}, data fetched from on-chip memory should 
be reused as much as possible before being discarded.

If the loops within an input feature map (Fig.~\ref{unrolling}(c)) are unrolled, each kernel weight 
is broadcast to multiply with P$ix\times$P$iy$ different input pixels in every cycle. Thus, it is 
reused P$ix\times$P$iy$ times. If multiple inputs in the batch are processed in parallel 
(Fig.~\ref{unrolling}(e)), the number of times the kernel weight is reused is equal to the batch 
size. If both types of loop unrolling are employed, the total number of times each kernel weight is 
reused is:
\begin{equation}
    weight\_reuse = Pix\times Piy\times batch\_size
\end{equation}

If loops across output feature map channels (Fig.~\ref{unrolling}(d)) are unrolled, then each input 
pixel is multiplied with multiple kernel weights from different kernels in parallel. Hence, each 
input pixel is reused P$of$ times. Besides, if both loops within a kernel window 
(Fig.~\ref{unrolling}(a)) and within an input feature map (Fig.~\ref{unrolling}(c)) are unrolled 
together, then the pixels in neighboring kernel windows partially overlap as long as the sliding 
stride is smaller than the kernel window size. This results in the average number of times each 
input pixel is reused being 
P$kx\times$P$ky\times$P$ix\times$P$iy$/(((P$ix-1$)S+P$kx$)((P$iy-1$)S+P$ky$)) since the overlapped 
pixels can be reused in the following cycle. Combining the three types of loop unrolling mentioned 
above results in the total number of times each input pixel is reused being \cite{Ma}:
\begin{equation}
    input\_reuse = \frac{Pof\times Pkx\times Pky\times Pix\times Piy}{((Pix-1)S+Pkx)((Piy-1)S+Pky)}
\end{equation}

\section{Analysis of DNN operation processing}\label{kernel}
In this section, we provide an analysis of hardware accelerator processing of DNN operations. We 
extend the analytical model for 2D convolution discussed in \cite{Ma} with batch processing enabled. 
Based on this approach, we build analytical models for various compute-intensive DNN operations. 

The number of MAC operations in the convolutional layer is 
N$_{MAC}$ =  N$if\times$N$kx\times$N$ky\times$N$ox\times$N$oy\times$N$of$. Ideally, the total 
number of cycles required is N$_{MAC}/$P$_{MAC}$, where P$_{MAC}$ is the total number of the MAC 
units and assuming 100\% MAC units efficiency. However, the available MAC units may not be fully 
utilized due to loop unrolling and loop tiling. In \cite{Ma}, the compute latency of the 
convolutional layer is modeled as the product of inter-tiling cycle and inner-tiling latency, where
\begin{equation}\label{inter}
\begin{multlined}
    inter\_tiling\_cycle =\\
    \lceil\frac{Nif}{Tif}\rceil\lceil\frac{Nkx}{Tkx}\rceil\lceil\frac{Nky}{Tky}\rceil\lceil\frac{Nox}{Tox}\rceil\lceil\frac{Noy}{Toy}\rceil\lceil\frac{Nof}{Tof}\rceil
\end{multlined}
\end{equation}

\begin{equation}
\begin{multlined}
    inner\_tiling\_latency =\\
    \lceil\frac{Tif}{Pif}\rceil\lceil\frac{Tkx}{Pkx}\rceil\lceil\frac{Tky}{Pky}\rceil\lceil\frac{Tox}{Pox}\rceil\lceil\frac{Toy}{Poy}\rceil\lceil\frac{Tof}{Pof}\rceil
\end{multlined}
\end{equation}
Inter-tiling cycle refers to the number of data chunks used in loop tiling and inner-tiling latency 
refers to the number of cycles required to process each chunk. Memory transfer latency is modeled 
as the maximum of input memory cycles and weight memory cycles, where
\begin{equation}
    num\_weight = Nox\times Noy\times Nkx\times Nky\times Nif\times Nof
\end{equation}
\begin{equation}
\begin{multlined}
    num\_input =\\
    Nox\times Noy\times Nkx\times Nky\times Nif\times Nof\times batch\_size
\end{multlined}
\end{equation}
\begin{equation}
    weight\_cycles = \lceil\frac{num\_weight}{weight\_reuse\times weight\_bandwidth}\rceil
\end{equation}
\begin{equation}
    input\_cycles = \lceil\frac{num\_input}{input\_reuse\times input\_bandwidth}\rceil
\end{equation}
With the assumption that memory bandwidth is not a bottleneck and multipliers can receive the input 
pixels and kernel weights continuously without incurring an idle cycle, the total processing latency 
for the convolutional layer is equal to the maximum value of the compute and memory transfer 
latencies.

To relax the constraint imposed by the memory bandwidth assumption made above and increase 
performance estimation accuracy, an extra optional finer-grained buffer simulator has been developed 
to monitor on-chip data. This on-chip buffer simulator estimates the memory transfer latency when the entire dynamic state of the DNN model cannot be held in the on-chip buffer. The entire convolutional layer is divided into multiple computational 
blocks that can be executed in parallel. Apart from the execution latency of each computational 
block, the memory transfer latency is also included if the data required by the block are not stored 
in the buffer. This buffer simulator simulates data fetching from and storing back to off-chip 
memory. The number of computational blocks represents a tradeoff between estimation speed and accuracy.

Depthwise separable convolution \cite{Xception} is a variation of 2D convolution. It
splits ordinary 2D convolution into two parts: 2D convolution within
each channel (depthwise convolution) and mixing the channels using a set
of 1$\times$1 convolutions across channels (channel mixing). Compared to ordinary convolution, it 
has fewer parameters. Therefore, it requires less computation and is less prone to overfitting. As 
shown in Table~\ref{parameters}, the first part, depthwise convolution, can be
fit into the 2D convolution model discussed above with the number of filter kernels being 
equal to 1. The second part, channel mixing, can be fit into the 2D convolution model with 
1$\times$1 kernel size. 

\begin{figure}[!t]
\centering
\includegraphics[width=3.5in]{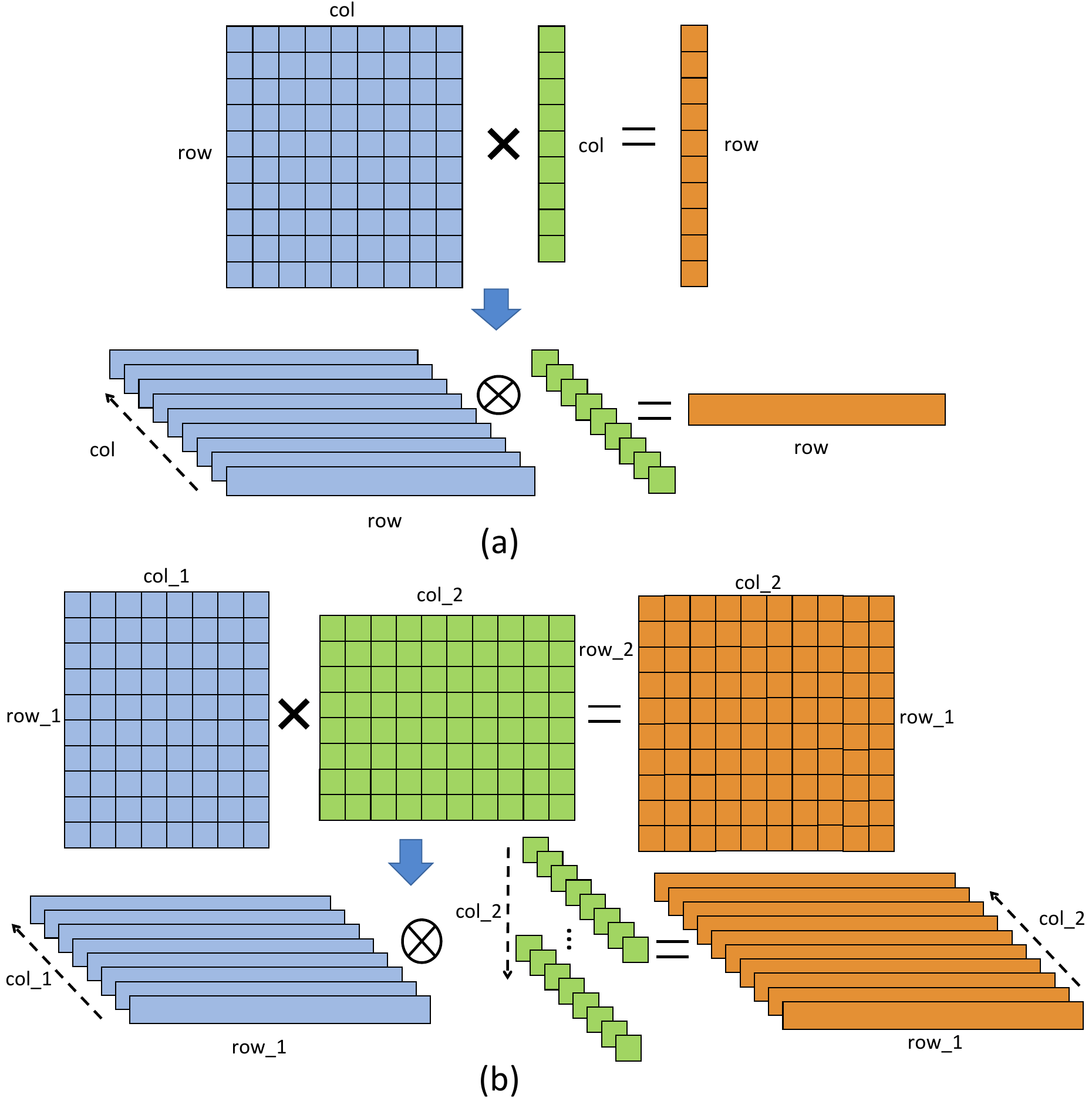}
\caption{Model transformations of matrix-vector and matrix-matrix multiplications}
\label{mul}
\end{figure}

Another important layer in a DNN is the fully-connected layer, which is processed as matrix-vector 
multiplication. We embed matrix-vector multiplication into 2D convolution to fit it into the 
analytical model described above, as shown in Fig.~\ref{mul}(a). The width, height, and depth of 
the input feature map are equal to the row number of the matrix, 1, and the column number of the matrix, 
respectively. The vector is transferred to the 1$\times$1 kernel with a depth equal to the matrix 
column number. Similarly, matrix-matrix multiplication is embedded into 2D convolution, as shown in 
Fig.~\ref{mul}(b). The second matrix is transferred to $col\_2$ 1$\times$1 kernels, where $col\_2$ 
is the column number of the second matrix. Details of the design parameter values used to fit 
depthwise separable convolution (depthwise convolution and channel mixing), matrix-vector 
multiplication, and matrix-matrix multiplication operations into the 2D convolution cost model are 
shown in Table~\ref{parameters}. In matrix-vector multiplication, col and row depict the matrix 
column number and row number, respectively. In matrix-matrix multiplication, col\_1, row\_1, and 
col\_2 depict the column and row numbers of the first matrix, and column number of the second matrix, 
respectively.

\begin{table}[t]
\centering
\caption{Operation parameter values used in the cost model}
\label{parameters}
\scalebox{0.75}{
\begin{tabular}{*{11}{|c}|}
\hline
2D conv. & Nif & Nix & Niy & Nkx & Nky & Nof & Nox & Noy & S\\
\hline
Depthwise conv. & Nif & Nix & Niy & Nkx & Nky & 1 & Nox & Noy & S\\
\hline
Channel mixing & Nif & Nix & Niy & 1 & 1 & Nof & Nox & Noy & S \\
\hline
Matrix-vector mul. & col. & row & 1 & 1 & 1 & 1 & row & 1 & 1\\
\hline
Matrix-matrix mul. & col\_1 & row\_1 & 1 & 1 & 1 & col\_2 & row\_1 & 1 & 1\\
\hline

\end{tabular}
}
\end{table}

\begin{table}[t]
\centering
\caption{Model validation with respect to an FPGA accelerator \cite{Ma}}
\label{validation}
\scalebox{0.95}{
\begin{tabular}{*{5}{|c}|}
\hline
\multirow{2}{*}{FPGA} & Frequency & Actual & Estimated & \multirow{2}{*}{Error} \\
& (MHz) & latency (ms) & latency (ms) & \\
\hline
Arria-10 GX 1150 & 150 & 47.97 & 38.73 & 19.3\% \\
\hline
\end{tabular}
}
\end{table}

We validated these DNN operation analytical models against the FPGA accelerator proposed in \cite{Ma} on VGG \cite{VGG}. As shown in Table~\ref{validation}, the error range is within 20\%.

\section{Application-driven architectural optimization}\label{framework}
In this section, we discuss the proposed application-driven architectural optimization framework that 
is based on the analytical operation models.

\subsection{Architectural optimization flow}
\begin{figure}[!t]
\centering
\includegraphics[width=3.5in]{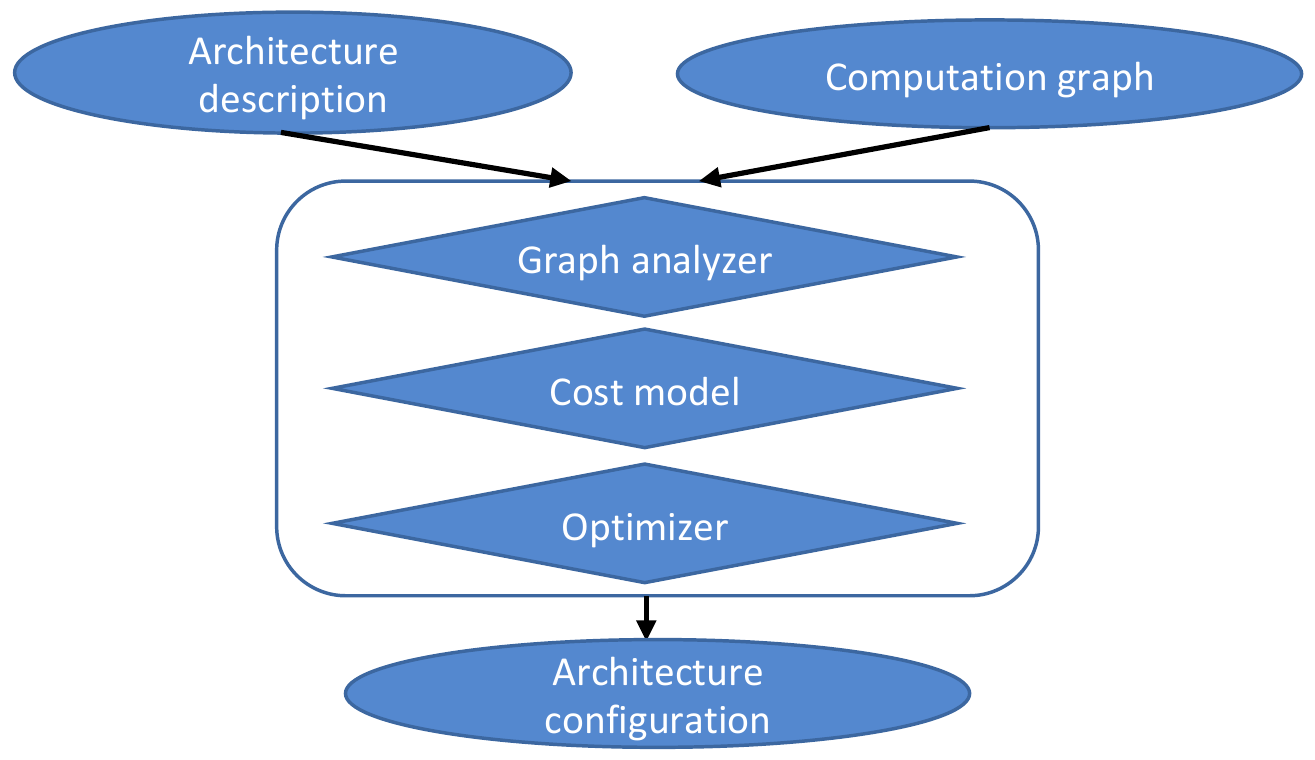}
\caption{Architectural optimization flow}
\label{flow}
\end{figure}

Fig.~\ref{flow} shows the accelerator architectural optimization flow. An architecture description 
file is used to define the design variables of the hardware accelerator. For example, it defines 
variables for the compute resource organization and the allocation of on-chip memory for activations 
and weights. Another input is the DNN computation graph of the target application that the accelerator 
is optimized for. We obtain this DNN computation graph by parsing the model file frozen from 
TensorFlow \cite{tensorflow}. It is a directed acyclic graph (DAG) in which a vertex represents a 
DNN operation and an edge defines data dependency. The computation graph is first analyzed by a 
graph analyzer to generate a DNN operation stream. Then the latency of each operation in the stream is estimated using the cost model  discussed in Section~\ref{kernel}. The total latency of the DNN model is estimated as the sum of these latencies. We only focus on the 
time-consuming operations. Accelerator performance on the target application is then optimized using 
a multidimensional optimizer to obtain an optimized architectural configuration. The graph analyzer and the multidimensional optimizer are discussed in the following sections.

\subsection{Computation graph analyzer}
The graph analyzer is used to transfer the model graph into a stream of operations, where the model execution is assumed to follow the order of the stream and the latency of the operations can be estimated using the cost models discussed in Section~\ref{kernel}. The DNN DAG is analyzed by performing depth-first search backward from the end node. The 
operation stream is obtained such that an operation can only be appended to the stream if it has no 
parent node or all of its parent nodes are already processed and are in the stream. The dynamic memory demand of the model is monitored during DAG traversal.

\begin{figure}[!t]
\centering
\includegraphics[width=3.5in]{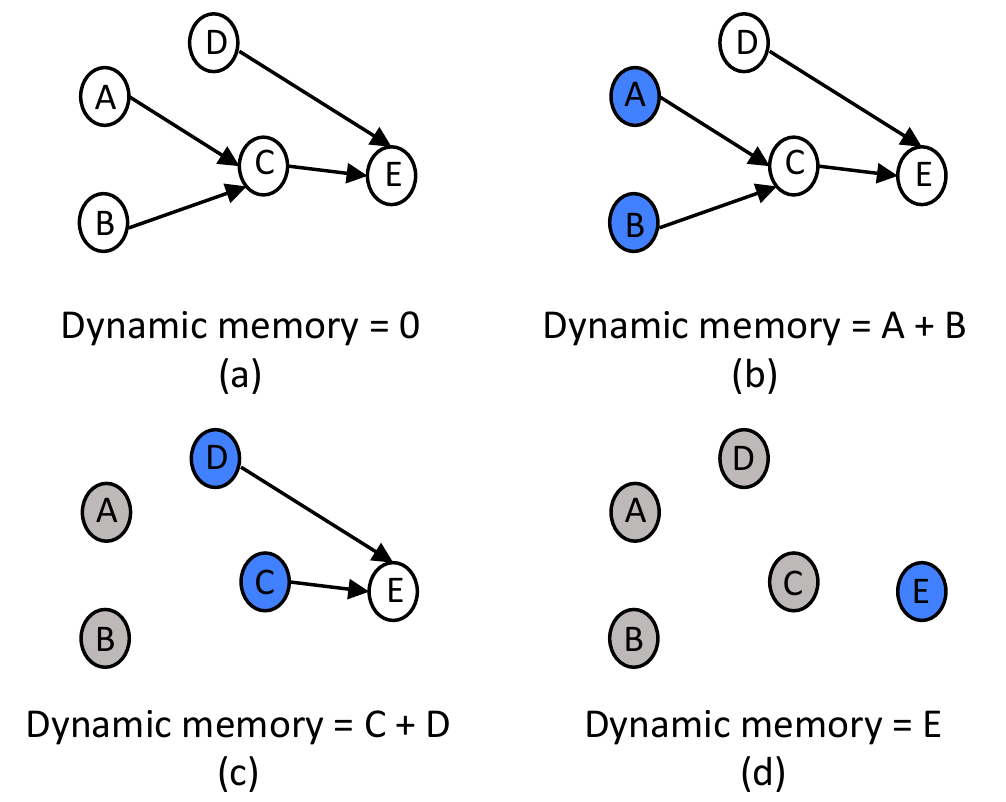}
\caption{Illustration of dynamic memory allocation analysis}
\label{dynamic}
\end{figure}

Fig.~\ref{dynamic}(a)-(d) show an example of a DAG and dynamic memory allocation analysis of the 
intermediate results. White nodes represent unprocessed operations that can only be processed if 
they have no parent nodes or all of their parent nodes have been processed. Then they become blue 
nodes whose outputs are stored in the on-chip memory. Their incoming edges are then removed since 
data dependency no longer exists after they are processed. A blue node turns to grey if it has no 
more outgoing edges, which means no more nodes depend on it. Hence, the memory space for its outputs 
can be deallocated. Dynamic memory allocation is monitored throughout DAG traversal and the maximum 
dynamic memory demand sets the lower bound for the on-chip buffer size of the accelerator if all the intermediate outputs (activations) need to be stored in the on-chip buffer.

\begin{table}[t]
\centering
\caption{Accelerator architecture design variables}
\label{variables}
\scalebox{0.75}{
\begin{tabular}{*{2}{|c}|}
\hline
Variable name & Definition\\
\hline \hline
loop\_order & The execution order of the convolutional loops\\
\hline
PE\_group & The total number of processing-element (PE) groups\\
\hline
MAC/group & The number of multiply-accumulates (MACs) in each PE group\\
\hline
buffer\_bank\_height & The height of the buffer bank\\
\hline
buffer\_bank\_width & The width of the buffer bank\\
\hline
weight\_bank/group & The number of buffer banks per PE group for weights\\
\hline
activation\_bank/group & The number of buffer banks per PE group for activations\\
\hline
Tif & The number of input feature map channels in loop tiling\\
\hline
Tix & The width of input feature map in loop tiling\\
\hline
Tiy & The height of input feature map in loop tiling\\
\hline
Tof & The number of output feature map channels in loop tiling\\
\hline
\end{tabular}
}
\end{table}

\subsection{Multidimensional optimization}\label{optimization}
We model the architectural optimization task as a multidimensional optimization problem. We choose 
performance under an area constraint as our optimization metric, where the DNN processing latency 
is estimated from the analytical model described above. The design variables defined in the 
architecture description file are the independent variables for multidimensional optimization. 
Table~\ref{variables} shows some of these variables. The minimum number 
of MAC units is constrained by the required number of parallel MAC operations per cycle:
\begin{equation}
\begin{multlined}
    PE\_group\times MAC/group \geq\\
    Pox\times Poy\times Pkx\times Pky\times Pif\times Pof
\end{multlined}
\end{equation}
The weight buffer size needs to be large enough to hold weight tiles. The maximum dynamic weight 
demand obtained from the computation graph analyzer sets the lower bound:
\begin{equation}
    weight\_buffer \geq Tkx\times Tky\times Tif\times Tof\times bit\_width 
\end{equation}
\begin{equation}
    weight\_buffer \geq peak\_weight\_memory\_demand
\end{equation}
Similarly, the constraints on the activation buffer size are:
\begin{equation}
\label{eq}
\begin{multlined}
    activation\_buffer \geq\\
    (Tix\times Tiy\times Tif+ Tox\times Toy\times Tof)\times bit\_width 
\end{multlined}
\end{equation}
\begin{equation}
\begin{multlined}
    activation\_buffer \geq\\ peak\_input\_memory\_demand\times batch\_size
\end{multlined}
\end{equation}

The $Tix\times Tiy\times Tif$ and $Tox\times Toy\times Tof$ products in Eq.~\ref{eq} correspond to 
input feature map tiling and output feature map tiling, respectively.

The accelerator area is estimated under the assumption of unit area for each 
component, e.g., MAC, control logic, on-chip buffer, and register file. The 
total area is then scaled according to the architectural configuration.

We use a genetic algorithm \cite{ga} to solve the multidimensional optimization problem. Other optimization methods, such as integer linear programming, may also be used. The pseudocode of the genetic algorithm is 
shown in Algorithm~\ref{GA}, where $K$, $\alpha$, $\beta$, and $\gamma$ denote the total number of generations, the percentage of top configurations that will pass to the next generation, the range of top configurations in the current generation that will be selected as parents in the crossover operation, and the probability of mutation, respectively. The performance of an accelerator architecture configuration is used as the fitness score. During crossover, a list of configuration variables is randomly selected. The offspring are then generated by swapping the variable values inherited from the parents. During mutation, a list of configuration variables is randomly selected. Then the given configuration variable values are subjected to mutation. We start with $n$ random initial valid accelerator configurations in the first generation. In each iteration, we first sort the configurations based on their performance. The top $\alpha n$ configurations pass through to the next generation. Then, the remaining configurations are generated through crossover with parents from the top $\beta n$ configurations in the current generation. Finally, $\gamma n$ configurations go through the mutation process to increase diversity. This process repeats until the total number of generations is reached or the performance improvement converges.

\begin{algorithm}[t]
\caption{Genetic algorithm}\label{GA}
\begin{algorithmic}[1]
\State Start with $n$ random initial valid accelerator configurations as the initial generation $G_0$
\For{$i \gets 0$ to $K-1$}
    \State Sort $G_i$ based on their fitness scores
    \State $G_{i+1} \gets G_i[0:\alpha n]$
    \While{$size(G_{i+1}) < n$}
        \State Randomly pick two numbers $f$ and $m$ in (0, $\beta n$)
        \State $F \gets G_i[f]$
        \State $M \gets G_i[m]$
        \State $G_{i+1} \gets G_{i+1} + Crossover(F, M)$
    \EndWhile
    \State Randomly select $\gamma n$ configurations $G_m$ in $G_{i+1}$
    \State $G_m \gets Mutation(G_m)$
\EndFor
\end{algorithmic}
\end{algorithm}

\section{Hardware-software co-design study}\label{results}
In this section, we study the relationship between the accelerator architecture and the 
characteristics of its target DNN applications based on the optimization framework.
We first optimize accelerator performance under an area constraint on the target applications 
through accelerator design space exploration. We provide an analysis of the characteristics of 
the different optimized architectures. We then explore the optimization opportunities for the 
accelerator architecture when multiple diverse DNN applications run simultaneously.
Finally, we study the relationships between DNN applications and the resulting optimized hardware 
accelerators.

\subsection{Accelerator architecture design space exploration}
We have selected eight representative DNNs: Inception-v3 (inception) \cite{inception}, 
DeepLabv3 (deeplab) \cite{deeplab}, ResNet-v1-50 (resnet) \cite{resnet}, Faster R-CNN (fasterRCNN) 
\cite{fasterrcnn}, PTB (ptb) \cite{ptb}, Wide \& Deep Learning (wdl) \cite{wdl}, NASNet 
(nasnet) \cite{nasnet}, and VGG16 (vgg) \cite{VGG}, and use the application-driven architectural optimization framework 
discussed in Section~\ref{framework} to optimize the accelerator performance under an area constraint.

Inception-v3 is a convolutional neural network (CNN) that uses filters with multiple sizes in the 
same layer. Extra 1$\times$1 convolutions are added before 3$\times$3 and 5$\times$5 convolutions to 
reduce the number of input feature channels, and thus the computational complexity of the network. 
DeepLabv3 is a CNN aimed at semantic image segmentation. It assigns labels to every pixel of the 
input image. It is constructed based on ResNet-101 \cite{resnet} and employs atrous spatial pyramid 
pooling for object segmentation at multiple scales. ResNet-v1-50 is a CNN that uses an ``identity 
shortcut connection" to solve the vanishing gradient problem \cite{VGP} during training. Faster 
R-CNN uses two networks for real-time object detection: a region proposal network for object boundary 
predictions and another network to detect objects in the bounding boxes. PTB is a recurrent 
neural network that uses LSTM units for word prediction.  Wide \& Deep Learning is 
a model for recommender systems. It jointly trains a wide linear model and a DNN for memorization and 
generalization, respectively.  NASNet is a network that is automatically generated by AutoML, which 
automates the design of machine learning models. It searches for the best layers on CIFAR-10 
\cite{cifar} and transfers the architectures to ImageNet \cite{ILSVRC15} for object detection. VGG16 is a convolutional neural network that replaces large kernel-sized filters in AlexNet with 3$\times$3 filters and stacks multiple convolutional layers on top of each other to increase depth and improve image recognition accuracy.

\begin{figure}[!t]
\centering
\includegraphics[width=3.5in]{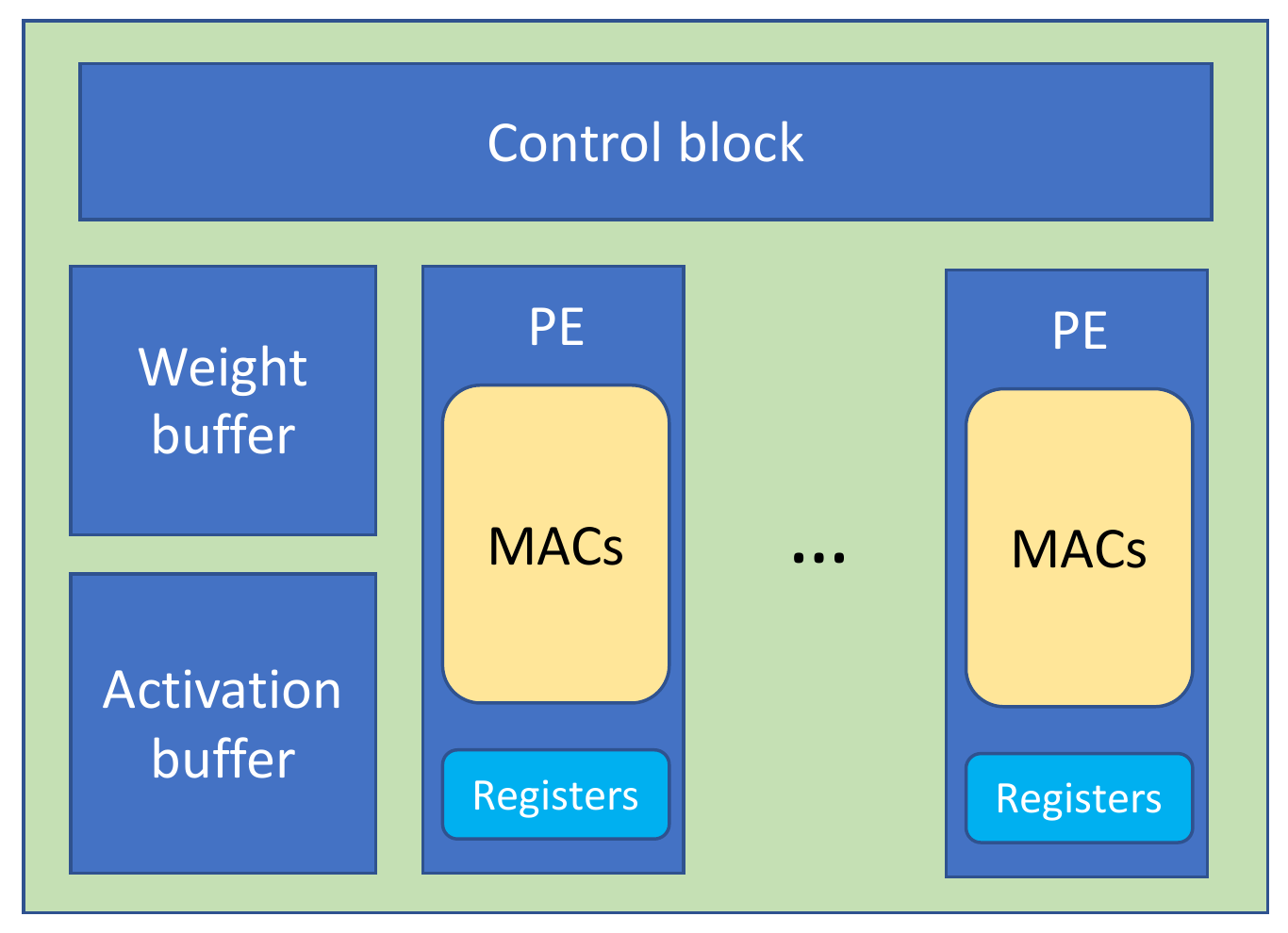}
\caption{Architectural template of the accelerator}
\label{template}
\end{figure}

\begin{figure*}[!t]
\centering
\includegraphics[width=7in]{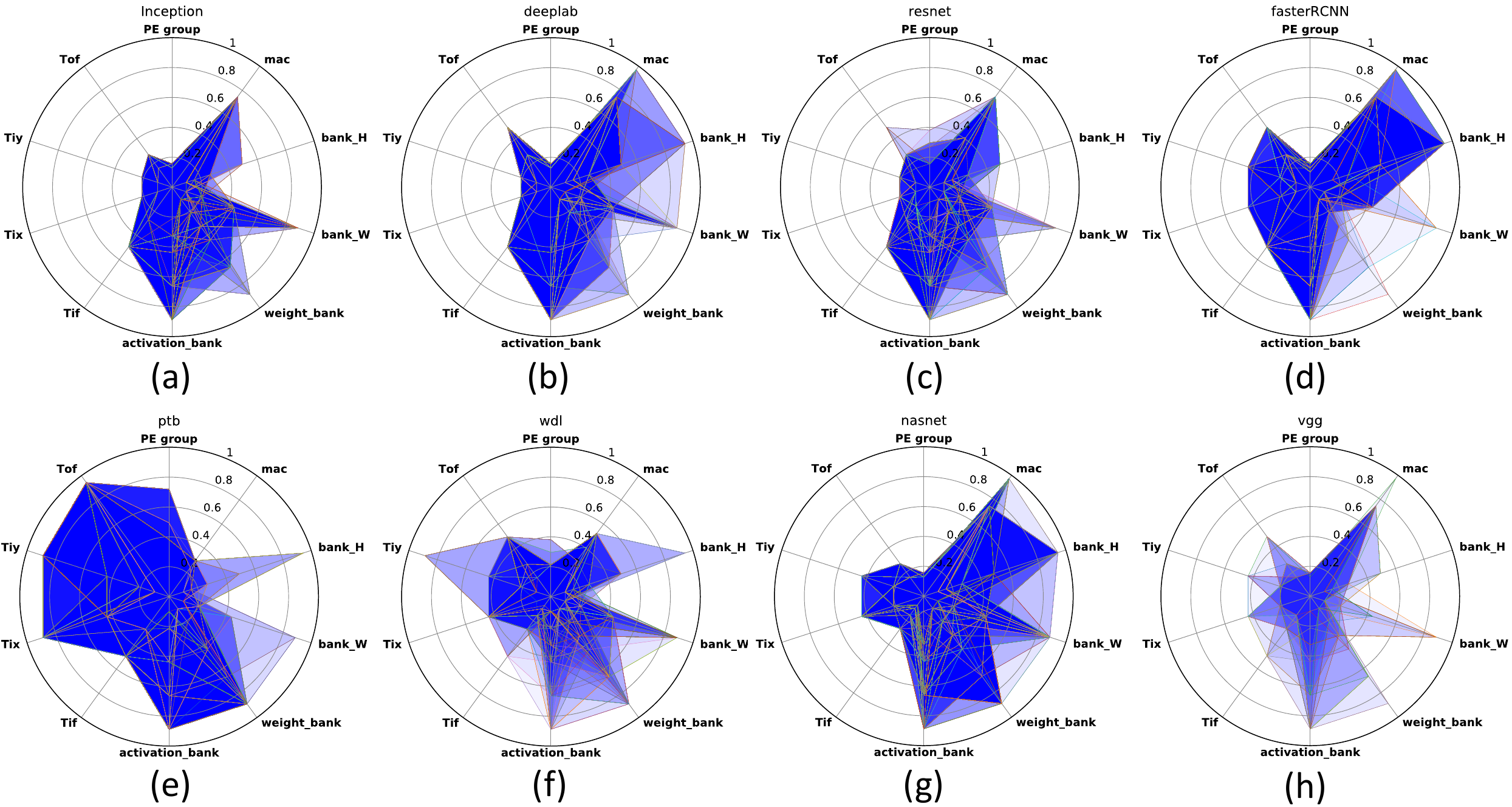}
\caption{Radar charts of accelerator configurations with top 10\% performance}
\label{radar}
\end{figure*}

\begin{figure*}[!t]
\centering
\includegraphics[width=7in]{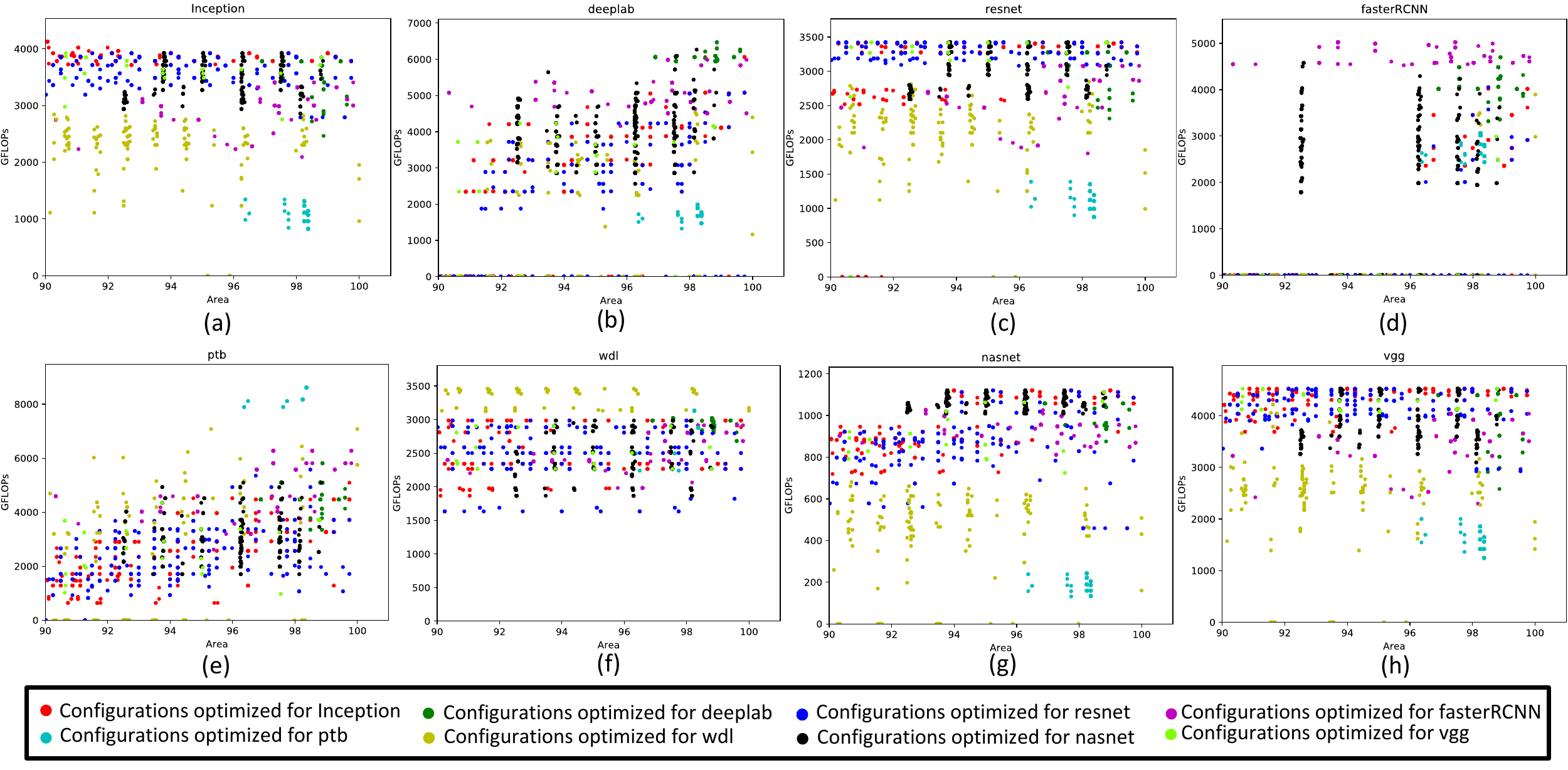}
\caption{Performance on selected DNN applications of accelerator configurations with top 10\% performance}
\label{plot}
\end{figure*}

Fig.~\ref{template} shows a template of the accelerator architecture. Two on-chip buffers are used to store weights and activations, respectively, and MACs are distributed in multiple PE groups. We assume that the accelerator bit-width is 8 and that it uses a batch size of 4. We select the obtained architectural configurations with top 10\% performance (in giga operations per
second (GOPS)) for each 
DNN application as candidates for optimized configuration selection. Their design configurations are
normalized for each variable and plotted in Fig.~\ref{radar}(a)-(h).  Their performance on the eight 
DNN applications is shown in Fig.~\ref{plot}(a)-(h). The highest performance on each DNN is achieved
by the architectural configuration optimized for that application using the framework. A configuration 
with 0 GOPS in Fig.~\ref{plot} means that the architecture violates the constraints mentioned in 
Section~\ref{optimization} for that specific application.

\begin{table*}[t]
\centering
\caption{Summary of the selected DNNs}
\label{networks}
\scalebox{1.14}{
\begin{tabular}{*{9}{|c}|}
\hline
 & inception & deeplab & resnet & fasterRCNN & ptb & wdl & nasnet & vgg\\
\hline
peak input memory demand & 2.8MB & 12.7MB & 2.4MB & 30.1MB & 8.0MB & 20.0KB & 5.3MB & 3.1MB\\
\hline
peak weight memory demand & 2.1MB & 12.8MB & 2.4MB & 0.3MB & 2.0MB & 8.0KB & 0.2MB & 2.0MB\\
\hline
\#Conv2D layers & 95 & 38 & 53 & 33 & 0 & 0 & 196 & 13\\
\hline
\#Depthwise separable convolutions & 0 & 17 & 0 & 13 & 0 & 0 & 160 & 0\\
\hline
\#Matrix-matrix mul. layers & 0 & 0 & 0 & 4 & 41 & 3 & 1 & 3\\
\hline
\end{tabular}
}
\end{table*}

\begin{figure}[!t]
\centering
\includegraphics[width=3.5in]{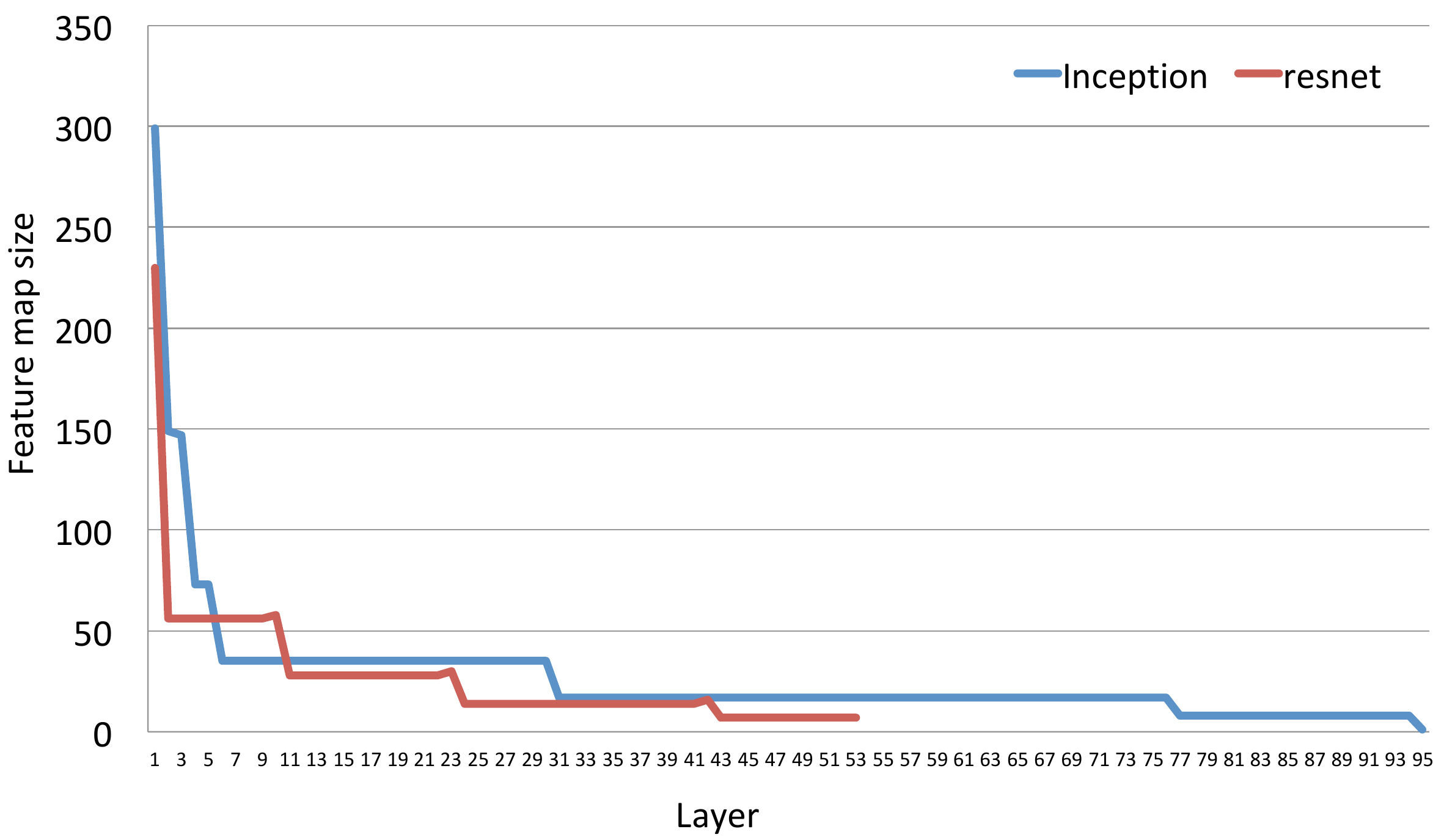}
\caption{Feature map size comparisons of Inception-v3 and ResNet-v1-50}
\label{size}
\end{figure}

\begin{figure}[!t]
\centering
\includegraphics[width=3.5in]{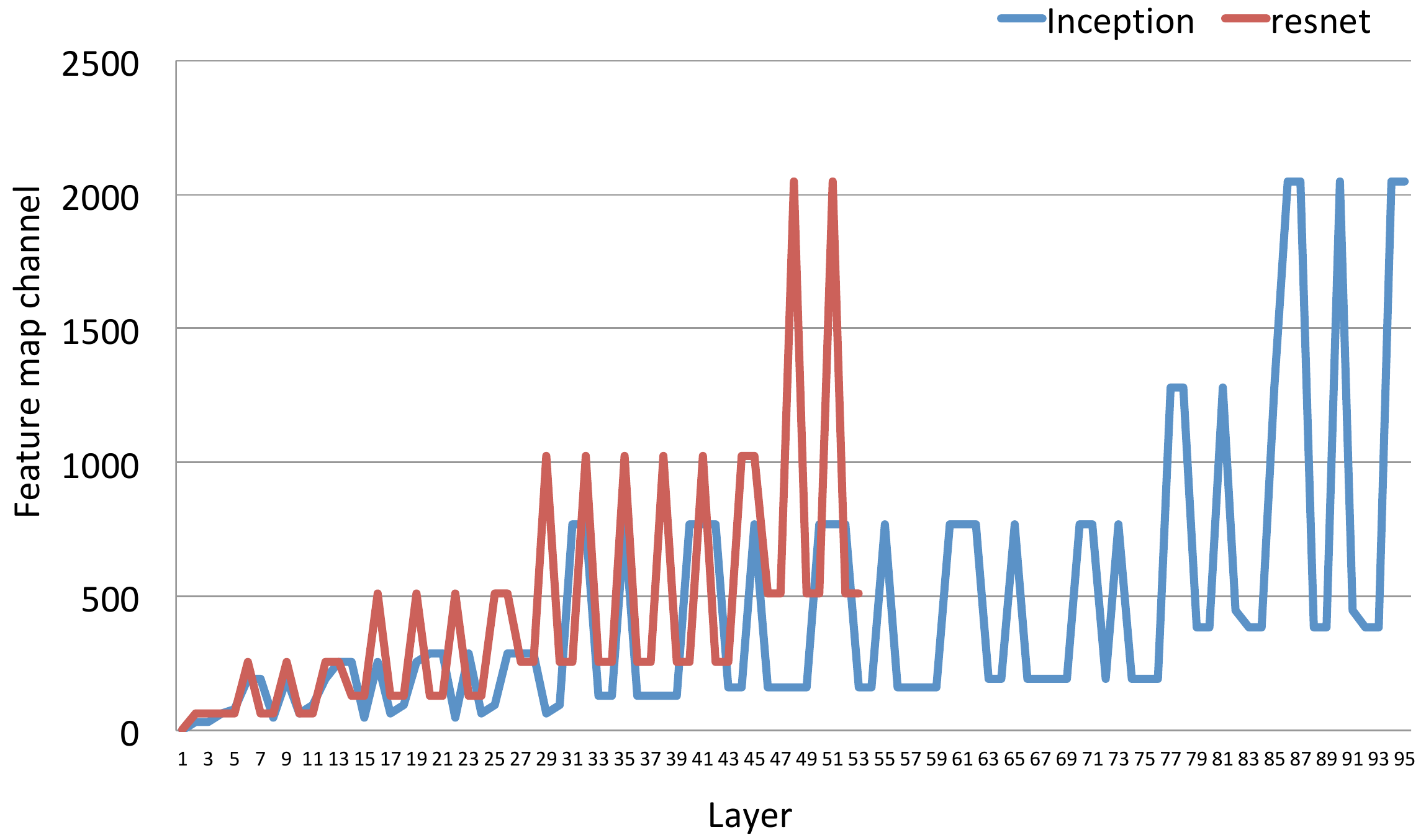}
\caption{Feature map channel comparisons of Inception-v3 and ResNet-v1-50}
\label{channel}
\end{figure}

We can see that Fig.~\ref{radar}(a) and Fig.~\ref{radar}(c) have similar
shapes. This means that the optimized architectures for Inception-v3
resemble those for ResNet-v1-50. This is consistent with the performance
plots in Fig.~\ref{plot}(a) and Fig.~\ref{plot}(c), respectively, where
they both achieve the highest performance on the two networks. The
reason for this architecture and resulting performance similarities is
that the two networks share similar characteristics, as shown in
Table~\ref{networks}. Inception-v3 and ResNet-v1-50 have similar peak
input/weight memory demands, which means that the two networks require
similar on-chip buffer size for the same data processing batch. This
is why Fig.~\ref{radar}(a) and Fig.~\ref{radar}(c) have the same values
for bank height, bank width, \#weight banks, and \#activation banks.
Besides, both networks mainly comprise 2D convolutional layers. Although
the depths of the two networks are different, the distributions of the
feature map size and the number of feature map channels are similar, as
shown in Fig.~\ref{size} and Fig.~\ref{channel}, respectively.
Architectures optimized for DeepLabv3 and Faster R-CNN also show
similarity in terms of their architectural configurations
(Fig.~\ref{radar}(b) and Fig.~\ref{radar}(d)) and performance on the two
networks (Fig.~\ref{plot}(b) and Fig.~\ref{plot}(d)). They both require
relatively larger on-chip memory for inputs. Therefore, there are dense
horizontal lines at 0 GOPS level in Fig.~\ref{plot}(b) and Fig.~\ref{plot}(d) 
because these architectural configurations violate on-chip memory constraints.

Among all candidate configurations, we select the one with the highest geometric 
mean of performance on the eight DNNs.
It is compared to the architectural configurations with the best
performance on each individual DNN, as shown in Table~\ref{geomean}. The
selected configuration outperforms the best configuration for each DNN
by 12.0\% to 117.9\% in terms of geometric mean performance, as shown in Table~\ref{gain}. 
The different characteristics of various DNNs may lead to significantly different 
configurations in the design space. Thus, the target applications should be considered 
in the early design stage to design efficient accelerators for a broad range of DNN applications. 

\begin{table*}[t]
\centering
\caption{Performance (GOPS) comparisons of the selected optimized configuration and the best 
configuration for each DNN. Performance on each DNN normalized to that of the best configuration}
\label{geomean}
\scalebox{1.1}{
\begin{tabular}{*{10}{|c}|}
\hline
& Best on & Best on & Best on & Best on & Best on & Best on & Best on & Best on & Selected\\
& inception & deeplab & resnet & fasterRCNN & ptb & wdl & nasnet & vgg & optimized result \\
\hline
inception & 1.00 &	0.73 & 0.98 & 0.76 &	0.27 & 0.59 & 0.99 & 0.98 & 0.94 \\
\hline
deeplab & 0.64 & 1.00 & 0.64 & 0.83 & 0.27 &  0.70 &	0.64 & 0.64 & 0.94\\
\hline
resnet & 0.99 &	0.76 & 1.00 & 0.77 & 0.32 & 0.61 & 0.99 & 0.99 & 0.96\\
\hline
fasterRCNN & 0.47 & 0.84 & 0.50 & 1.00 &	0.55 & 0.69 & 0.47 & 0.50 & 0.84 \\
\hline
ptb & 0.38 & 0.54 &	0.43 & 0.47 & 1.00 &	0.62 & 0.38 & 0.43 & 0.59\\
\hline
wdl & 0.68 & 0.85 &	0.66 & 0.69 &	0.83 & 1.00 & 0.68 &	0.66 & 0.83\\
\hline
nasnet & 0.99 &	0.86 &	0.99 &	0.89 &	0.17 & 0.46 & 1.00 & 0.99 &	0.94\\
\hline
vgg & 0.99 & 0.71 &	0.99 & 0.78 &	0.35 & 0.56 & 0.99 &	1.00 & 0.98\\
\hline
Geometric mean & 0.72 &	0.78 &	0.74 &	0.76 & 0.40 & 0.64 & 0.72 & 0.74 &	0.87\\
\hline
\end{tabular}
}
\end{table*}

\begin{table*}[t]
\centering
\caption{Average performance improvements of the selected result over the best configuration for 
each DNN}
\label{gain}
\scalebox{1.07}{
\begin{tabular}{*{8}{|c}|}
\hline
Over the best & Over the best & Over the best & Over the best & Over the best & Over the best & Over the best & Over the best\\
on inception & on deeplab & on resnet & on fasterRCNN & on ptb & on wdl & on nasnet & on vgg\\
\hline
20.0\% & 12.0\% & 18.0\% & 14.8\% & 117.9\% & 36.1\% & 20.0\% & 18.0\%\\
\hline
\end{tabular}
}
\end{table*}

\subsection{Multi-context optimization}
From Fig.~\ref{radar}, we observe that the optimized accelerator architectures diverge for 
different DNN applications. Hence, in this section, we explore if there exist new optimization 
opportunities when very different DNN applications run simultaneously on the same hardware accelerator.

\begin{figure}[!t]
\centering
\includegraphics[width=3.5in]{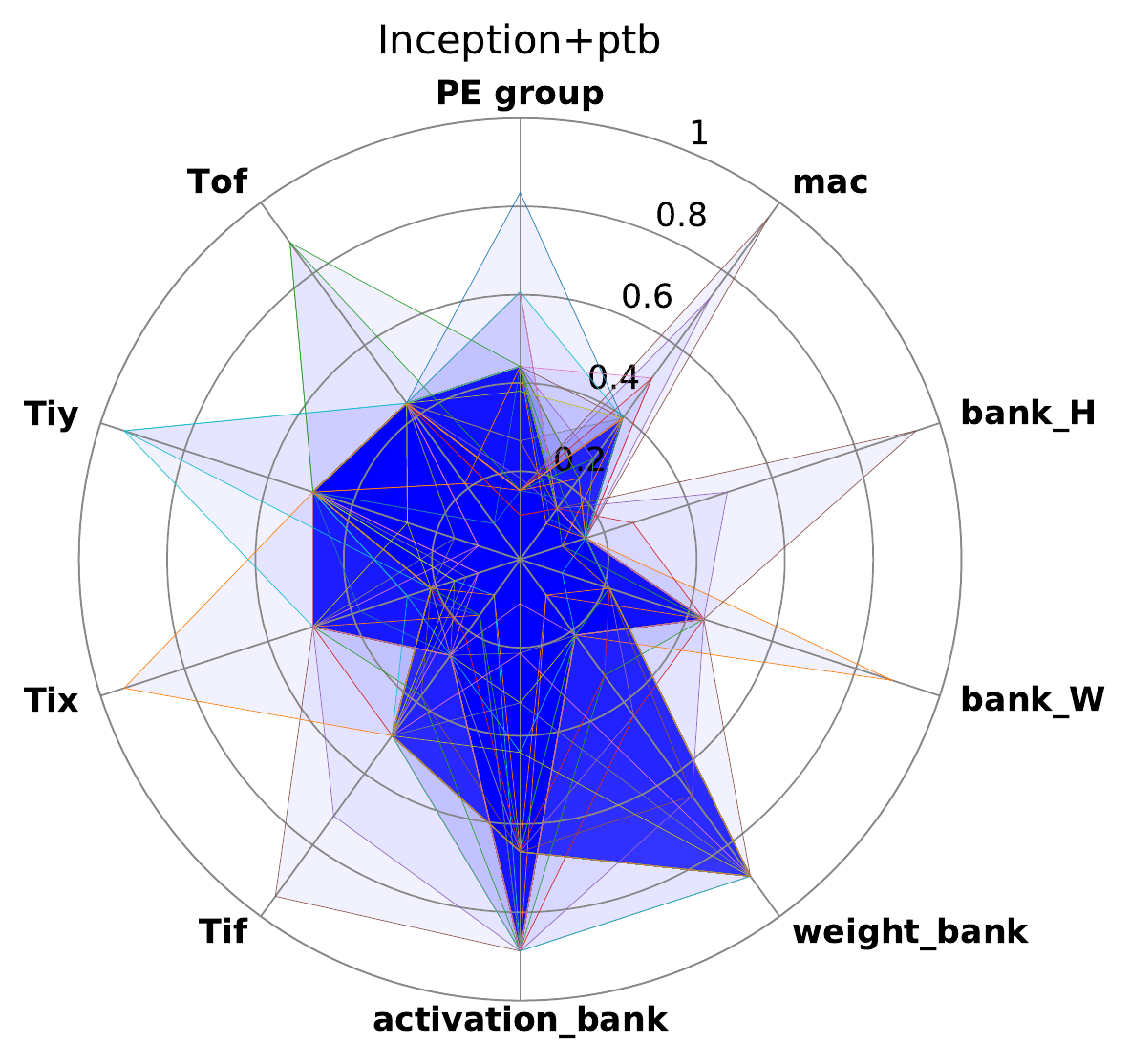}
\caption{Radar chart of accelerator configurations for a multi-context application}
\label{multicontext}
\end{figure}

First, we mix Inception-v3 and PTB by interleaving layers from both DNNs. Then, we use the framework 
to optimize the accelerator architecture on this mixed DNN in a multithreaded manner: the operations from Inception-v3 and PTB run on the accelerator alternately. The multithreading scenario is simplified using this mixed-layer stream. In real scenarios, a more sophisticated scheduler would be needed to decide the layer execution order for the two models. The inference results obtained using the multithreading mixed-layer stream are the same as when the two models are run back-to-back, as layers from different models are tagged and data are not shared across models.

Fig.~\ref{multicontext} shows the resulting architectural configurations with top 10\% performance 
on this multi-context application. This radar chart is quite different from those shown in 
Fig.~\ref{radar}(a) and Fig.~\ref{radar}(e) and is not a simple combination of those two radar 
charts.  It has smaller \#macs compared to Fig.~\ref{radar}(a) and smaller loop tiling sizes, e.g., 
$Tof$, $Tiy$, and $Tix$, relative to Fig.~\ref{radar}(e). As shown in Table~\ref{networks}, Inception-v3 
is compute-intensive: it is dominated by 2D convolutional layers and thus requires relatively 
larger \#macs for efficient processing. On the other hand, PTB is memory-intensive: it consists of 
a large number of matrix-matrix multiplication layers with relatively high peak input/weight demand. 
Hence, large tiling sizes appear in its optimized architectural configurations. However, when these 
two DNN applications run simultaneously on the same accelerator, the required amount of compute 
and memory resources is lowered in the optimized configurations. The reasons for this are two-fold. 
First, under an area constraint, the optimized architectural configurations for the multi-context 
application need to maintain a balance between compute and memory resources. Second, the 
complementary characteristics of Inception-v3 and PTB help relax both compute and memory design 
constraints on the accelerator architecture: while MACs are mainly devoted to convolutional layers 
of Inception-v3, filter weights can be transferred between the weight buffer and external memory for 
matrix-matrix multiplication layers for PTB at the same time, with no or very little performance 
loss, since the layers of both DNNs are interleaved. This shows that the optimal design for 
multi-context applications may not be a simple combination of designs optimized for the individual 
applications and new optimization opportunities can be explored using our application-driven 
architectural design space exploration framework.

\subsection{Application sensitivity analysis}
It is evident that the application-driven architectural optimization framework will generate similar 
architectural configurations for DNNs with common characteristics. However, to better understand the 
reasons for the different accelerator configuration results shown in Fig.~\ref{radar}, we perform an 
application sensitivity analysis to discover the hardware-software relationship in DNN applications.

We build the Faster R-CNN network in four steps.  In the first step, we build a DNN with the same 
number of 2D convolutional layers as that in Faster R-CNN, but with relatively larger feature map 
sizes. The next step is to make the convolution dimensions the same as those in the Faster R-CNN. 
Depthwise separable convolutional layers and matrix-matrix multiplication layers are then added in 
the following two steps. In each step, we use the architectural optimization framework to generate 
the architectural configurations and select those with top 10\% performance. 

\begin{figure}[!t]
\centering
\includegraphics[width=3.5in]{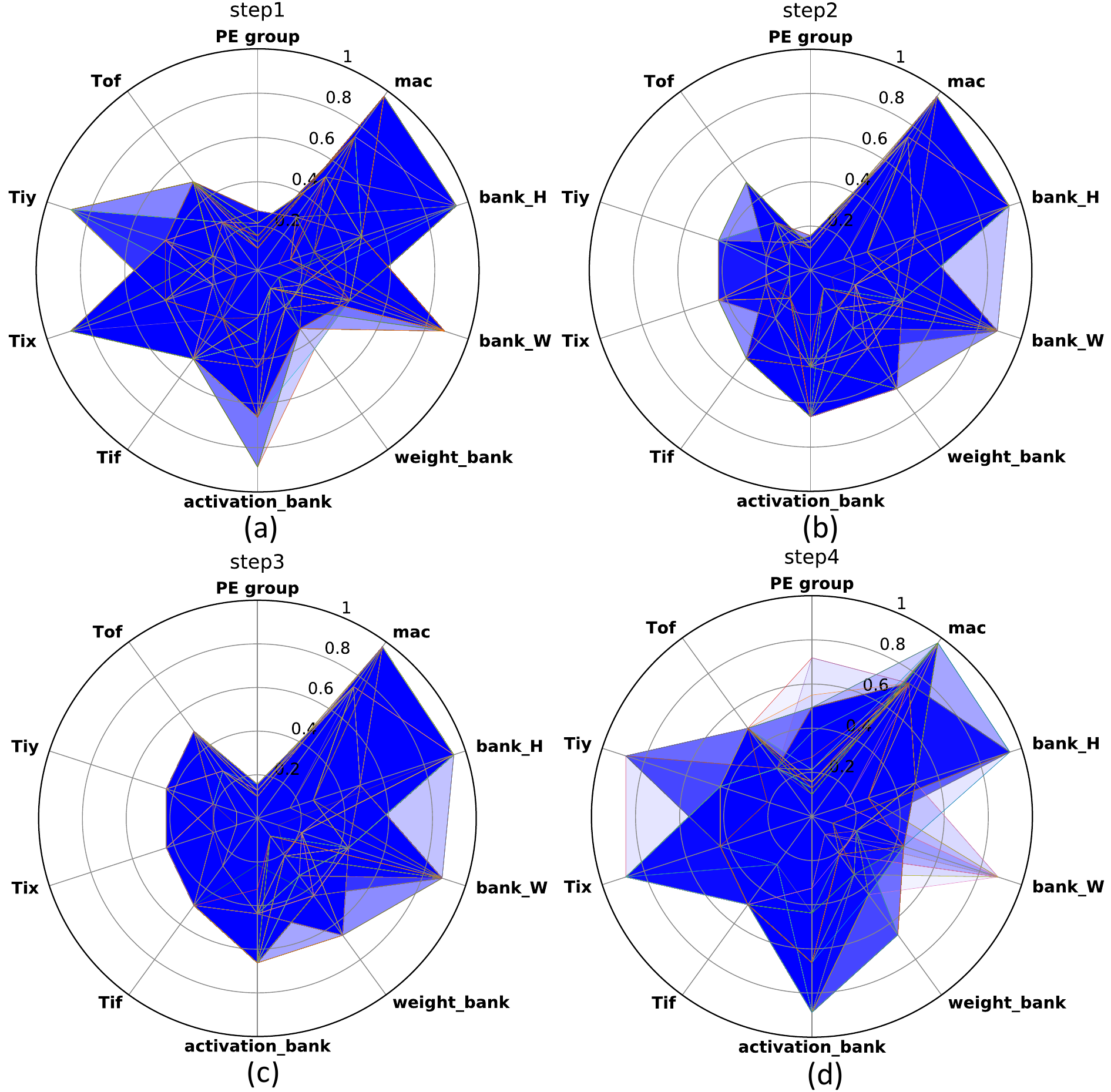}
\caption{Radar charts of accelerator configurations at each step}
\label{transfer}
\end{figure}

Fig.~\ref{transfer}(a)-(d) show the optimized architectural configurations obtained at each step. 
We can see that reducing the feature map sizes of the convolutional layers (from 
Fig.~\ref{transfer}(a) to Fig.~\ref{transfer}(b)) impacts the loop tiling design variables. Smaller 
tiling sizes are preferred for better performance. According to Eq.~(\ref{inter}), reducing the 
feature map size while keeping loop tiling variables unchanged may lower the efficiency of memory 
transactions. Thus, the value of loop tiling variables is also reduced in Fig.~\ref{transfer}(b). 
In the third step, 13 depthwise separable convolutional layers are inserted every two Conv2D layers 
with the same convolution dimensions as their following Conv2D layers. Comparing 
Fig.~\ref{transfer}(b) and Fig.~\ref{transfer}(c), we can see that just adding depthwise separable 
convolution operations without changing the feature map size does not affect the optimized 
architectural configurations. In the fourth step, large matrix multiplication layers are added. 
The number of PE groups is increased to generate more parallel MACs for matrix multiplication 
processing. Besides, this also increases the value of loop tiling variables again since more 
computational parallelism can be exploited when processing larger data chunks. This is consistent 
with the architectural configurations optimized for PTB, as shown in Fig.~\ref{radar}(e), where the 
network only consists of large matrix multiplication layers. Fig.~\ref{radar}(a)-(c) and 
Fig.~\ref{radar}(g)-(h) have small design variables in loop tiling dimensions since there are no matrix 
multiplication layers or the matrix dimension is small.

If the underlying hardware compute resource is fixed, we can perform a similar sensitivity analysis 
on the target network using this application-driven architectural optimization framework. The 
analysis results can guide DNN model development to fit the underlying compute resource.

\section{Discussions and limitations}\label{discussion}
In this section, we discuss the assumptions we made in our framework and identify future work to tackle these limitations.

Similar to the designs in \cite{Qiu,Han,chain,fuse}, we assume the architectural template uses dedicated separate weight and activation on-chip buffers. However, there are other designs in which a unified memory is used for weights and activations. A future direction for improving our design space exploration framework is to support a unified memory design.

Another limitation of our memory modeling technique is that it does not model memory distribution across multiple PEs, hence overlooking data sharing and replication. A more accurate memory modeling technique is required for this purpose.

Since our analytical modeling framework is useful for early stage design space exploration, it is very sensitive to modeling latency in order to explore the huge design space. A more accurate architecture/memory modeling method may be too slow for early design space exploration. Therefore, a hierarchical modeling method could be used to obtain a balance between modeling accuracy and latency: a coarse-grained modeling method in the first step, where a small number of candidate configurations are selected, and a finer-grained modeling method applied in the second step to those selected configurations to obtain the final results. We leave this as future work.

\section{Conclusion}\label{conclusion}
In this article, we proposed an application-driven accelerator architectural optimization framework. 
This framework explores the accelerator design space and optimizes the architectural configuration 
for the target applications based on the analytical models. We use a genetic algorithm to 
solve the multi-dimensional optimization problem. We use this framework to optimize the accelerator 
architectural configuration for eight selected DNN applications.  We show that the architectural 
configuration optimized for all the eight DNNs can achieve geometric mean performance improvements 
ranging from 12.0\% to 117.9\% over the configurations optimized only for each individual DNN.
In addition, we explore the opportunity to use the framework for accelerator architectural 
configuration optimization when complementary DNN applications run simultaneously.  Furthermore, the 
framework can be used to guide DNN model development for running on the fixed hardware accelerator 
more efficiently.

\ifCLASSOPTIONcaptionsoff
  \newpage
\fi

\bibliographystyle{IEEEtran}

\bibliography{bare_adv}

\begin{IEEEbiography}[{\includegraphics[width=1in,height=1.25in,clip,keepaspectratio]{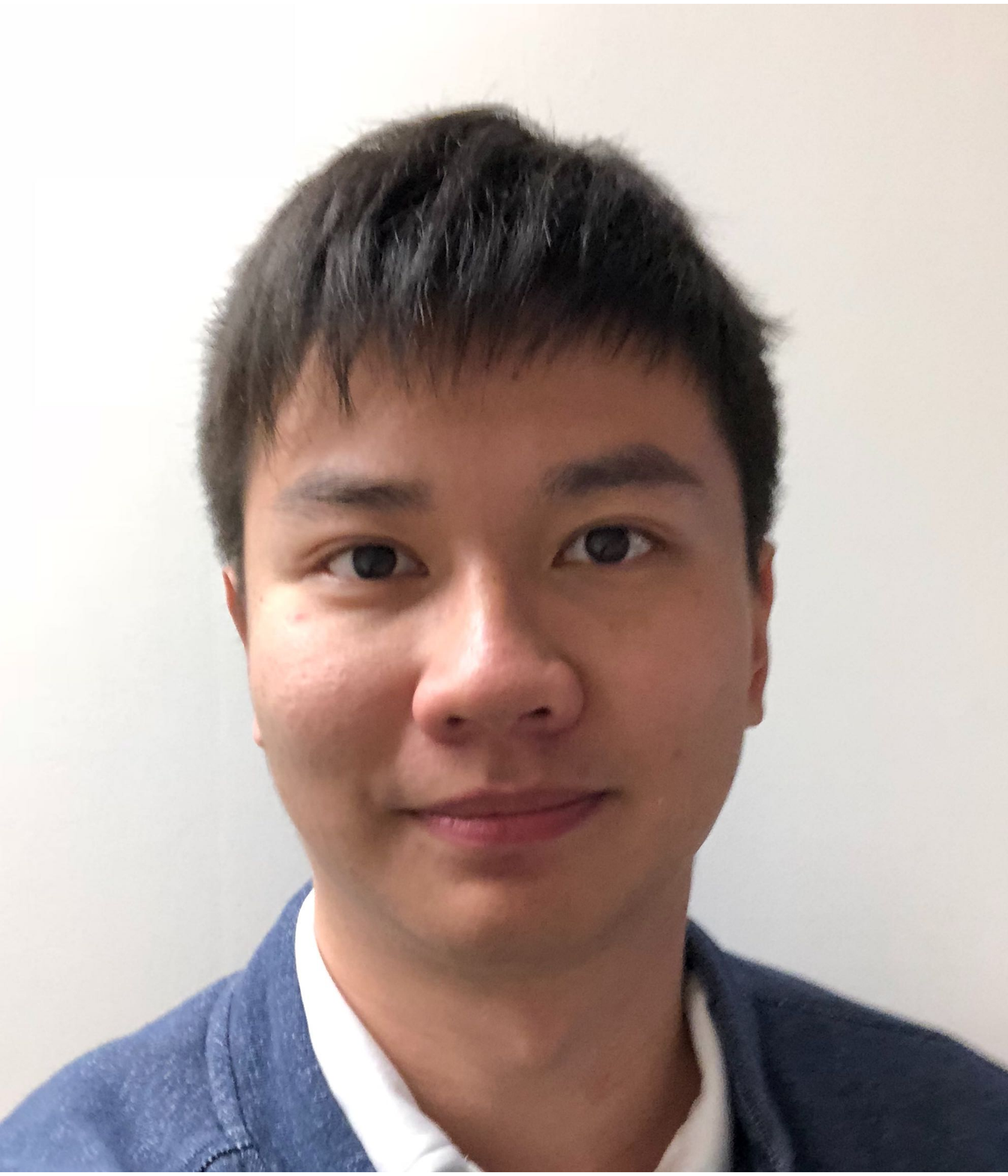}}]{Ye Yu}
received the B.Eng. degree in Electronic and Computer Engineering from The 
Hong Kong University of Science and Technology, Hong Kong, China, in 2014, 
and the M.A. and Ph.D. degrees in Electrical Engineering from Princeton University, NJ, 
USA, in 2016 and 2019, respectively. He is currently a software engineer at Microsoft.

His current research interests include computer vision, machine learning, deep learning model compression and acceleration.
\end{IEEEbiography}

\begin{IEEEbiography}[{\includegraphics[width=1in,height=1.25in,clip,keepaspectratio]{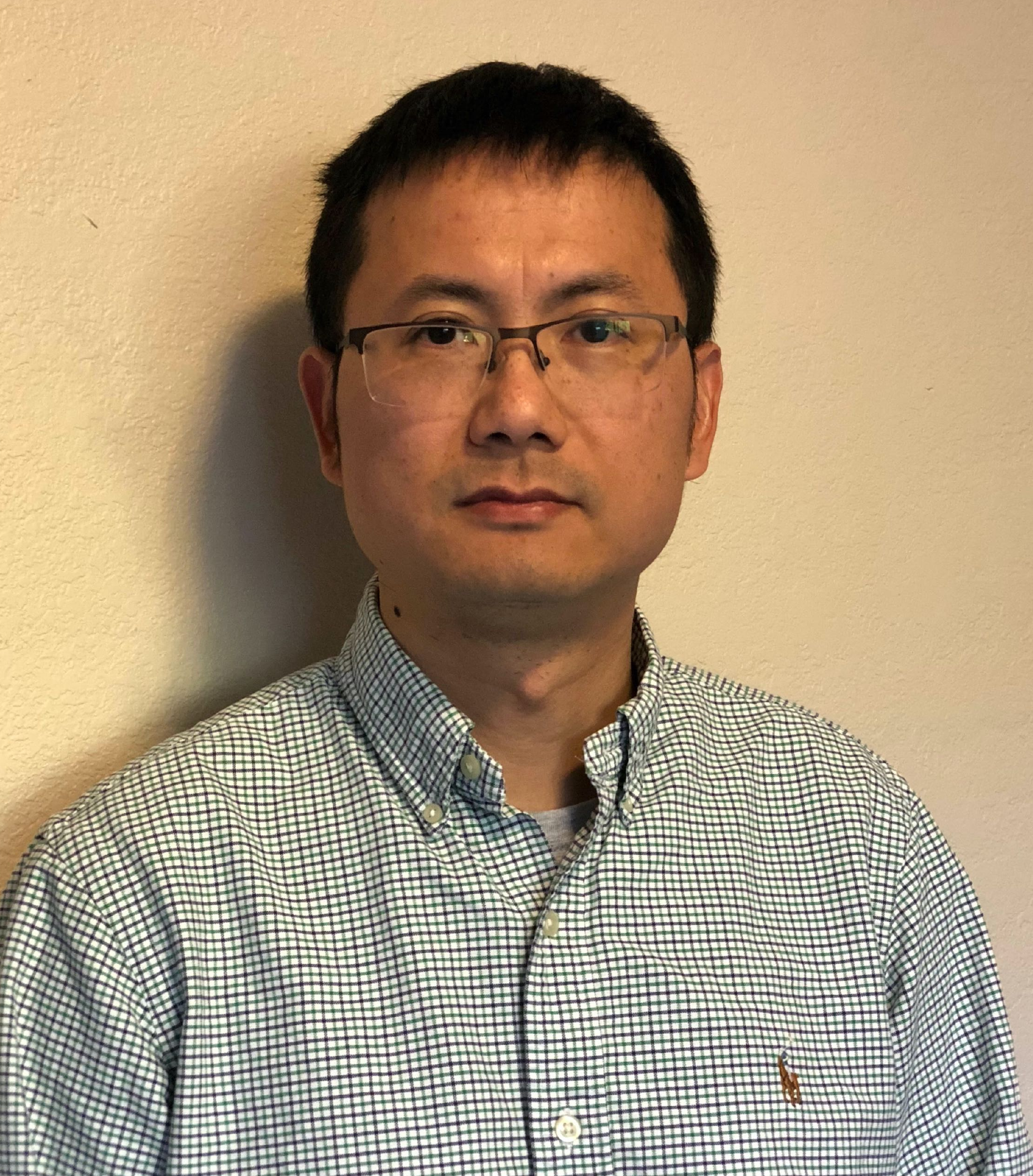}}]{Yingmin Li}
Yingmin Li received his B.S. in Computer Science from Zhejiang University, China and Ph.D. in 
Computer Science from University of Virginia. Yingmin is currently a senior staff engineer at 
Alibaba's Infrastructure Services Group. His research interests include machine learning, performance 
optimization for heterogeneous systems, and computer architecture.
\end{IEEEbiography}

\begin{IEEEbiography}[{\includegraphics[width=1in,height=1.25in,clip,keepaspectratio]{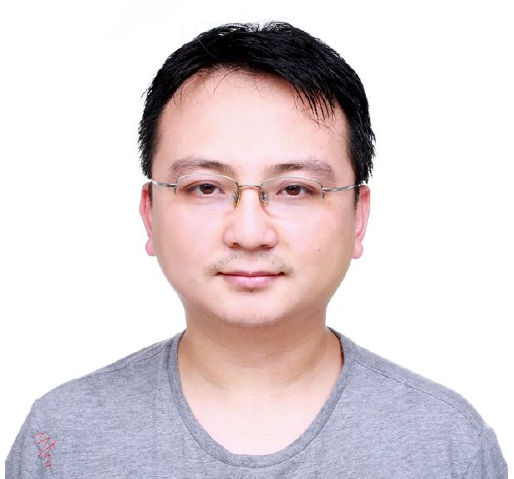}}]{Shuai Che}
Shuai Che received the Ph.D. degree in Computer Engineering from the University of Virginia in 2012 
and Bachelor of Engineering degree from Shanghai Jiaotong University in 2004.

His research interests include computer architecture, heterogeneous computing, GPGPU, machine 
learning, and graph computing. 
\end{IEEEbiography}

\begin{IEEEbiography}[{\includegraphics[width=1in,height=1.25in,clip,keepaspectratio]{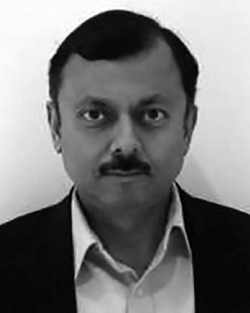}}]
{Niraj K. Jha} (S'85-M'85-SM'93-F'98) received his B.Tech. degree in Electronics and
Electrical Communication Engineering from Indian Institute of Technology,
Kharagpur, India in 1981, M.S. degree in Electrical Engineering from S.U.N.Y.
at Stony Brook, NY in 1982, and Ph.D. degree in Electrical Engineering from
University of Illinois at Urbana-Champaign, IL in 1985. He is a Professor of
Electrical Engineering at Princeton University. 

He has served as the Editor-in-Chief of IEEE Transactions on VLSI Systems and 
an Associate Editor of IEEE Transactions on Circuits and Systems I and II, 
IEEE Transactions on VLSI Systems, IEEE Transactions on Computer-Aided Design, 
IEEE Transactions on Computers, IEEE Transactions on Multi-Scale Computing System, 
Journal of Electronic Testing: Theory and Applications, and Journal of Nanotechnology. 
He is currently serving as an Associate Editor of Journal of Low Power Electronics. He has 
served as the Program Chairman of several conferences, the Director
of the Center for Embedded System-on-a-chip Design funded by New Jersey
Commission on Science and Technology, and the Associate Director of the
Andlinger Center for Energy and the Environment.

He is the recipient of the AT\&T Foundation Award and NEC Preceptorship Award 
for research excellence, NCR Award for teaching excellence, Princeton 
University Graduate Mentoring Award, and six Dean's Teaching Commendations
from the School of Engineering and Applied Sciences.  He is a Fellow of IEEE 
and ACM. He received the Distinguished Alumnus Award from I.I.T., Kharagpur in 2014.

He has co-authored five books, 15 book chapters, and more than 440 technical papers, 
14 of which have won various awards and six more that have been nominated for 
best paper awards.  He has received 17 U.S. patents.

His research interests include monolithic 3D IC design, low power 
hardware/software design, computer-aided design of integrated circuits and 
systems, machine learning, smart healthcare, and secure computing. He has 
given several keynote speeches in the area of nanoelectronic design/test
and smart healthcare.
\end{IEEEbiography}

\begin{IEEEbiography}[{\includegraphics[width=1in,height=1.25in,clip,keepaspectratio]{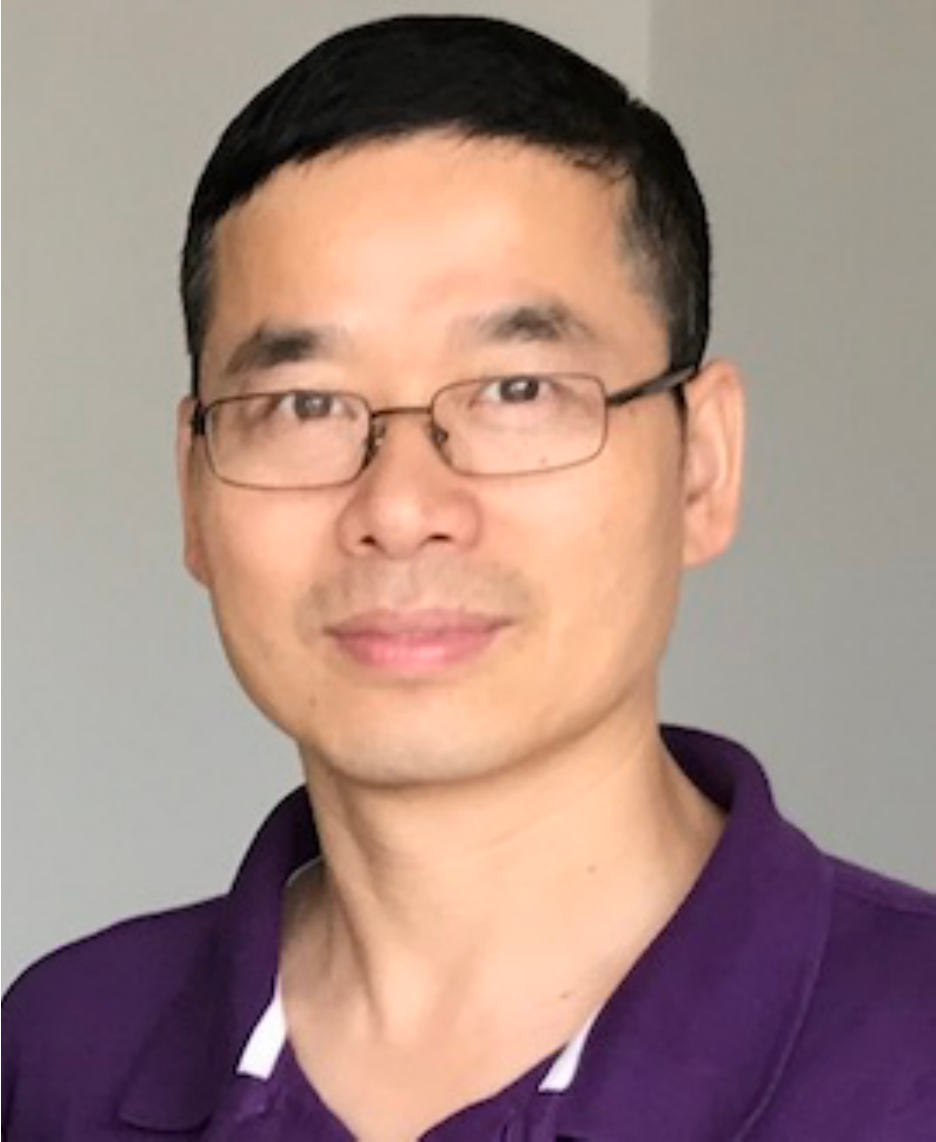}}]{Weifeng Zhang}
Dr. Weifeng Zhang received his B.Sc. from Wuhan University, China and PhD in Computer Science \& 
Engineering from University of California San Diego in 2006. Weifeng currently is the Chief Scientist 
of Heterogeneous Computing at Alibaba Infrastructure Services Group. His research interests include 
computer architecture, compilation, architectural support for dynamic optimization, machine learning, 
and acceleration via heterogeneous computing architectures.
\end{IEEEbiography}

\end{document}